\documentclass[manuscript]{acmart}
\renewcommand\footnotetextcopyrightpermission[1]{} 
\AtBeginDocument{%
  \providecommand\BibTeX{{%
    \normalfont B\kern-0.5em{\scshape i\kern-0.25em b}\kern-0.8em\TeX}}}

\acmBooktitle{The CHI 23 Workshop on Combating Toxicity, Harassment, and Abuse in Online Social Spaces,
 April 23--28, 2023, Hamburg, Germany} 

\begin{document}

\title{Toxicity and Cultural Entrenchment in Peer-Production Communities: Toward a Handbook on Intelligent System Design}
\renewcommand{\shorttitle}{Toxicity and Cultural Entrenchment in Peer-Production Communities}

\author{Chris Blakely}
\email{c_blakely@kcg.edu}
\orcid{0002-9503-9797}
\affiliation{%
  \institution{The Kyoto College of Graduate Studies for Informatics}
  \city{Kyoto}
  \country{Japan}
  \postcode{}
}

\author{Andrew Vargo}
\email{awv@omu.ac.jp}
\orcid{}
\affiliation{%
  \institution{Osaka Metropolitan University}
  \city{Sakai}
  \country{Japan}
}

\renewcommand{\shortauthors}{Blakely \& Vargo.}

\begin{abstract}
Toxicity and abuse are common in online peer-production communities. The social structure of peer-production communities that aim to produce accurate and trustworthy information require some conflict and gate-keeping to spur content production and curation. However, conflict and gate-keeping often devolve into hierarchical power structures which punish newcomers and lock out marginalized groups through entrenched cultural norms. Community administrators often focus on content quality, rather than consideration for all user safety, to promote community growth and survival. Once toxic cultural norms dominate a peer-production community, it is very difficult for community administrators to stop these behaviors from undermining inclusive peer-production. We propose developing a "handbook of intelligent system design" that attempts to frame design protocols to better read user-community culture and accurately distinguish toxic negative interactions from beneficial conflict. 

\end{abstract}




\maketitle
\thispagestyle{empty} 

\section{Introduction: Social Conflict and Toxic Conflict}
Conflict\footnote{This manuscript was submitted to The CHI 23 Workshop on Combating Toxicity, Harassment, and Abuse in Online Social Spaces,
 April 23--28, 2023, Hamburg, Germany} is often conflated with toxicity and can sometimes lead to abusive user behavior, but not all conflict or negative interactions are or will become toxic~\cite{cheng2017anyone,koehler2018really}. Social conflict and tribalism develop for various reasons irrespective of user perceptions of harassment~\cite{ouvrein2018online, campbell2009conflict}. Since some form of conflict in peer-production communities is necessary to stimulate discussion, some tension is important to the community culture. There is trouble however, when the community becomes so entrenched as to not allow newcomers to join the community. This can then lead to a situation in which marginalized groups are effectively banned from full participation in the system. The community may try to mitigate this behavior with campaigns and new rules, but this will often fail~\cite{vargo2016corrective}.

We focus on peer-production communities, which we define as communities where users interact with each other to produce valuable content for a specific goal or interest, because these communities play a major role in online socialization and professional development. In these communities, conflicts occur because of 1. debate/discussion, 2. user correction, or 3. failure to follow community standards~\cite{vargo2018identity}. In this position paper, we assume that the peer-production system seeks to be fully inclusive.

\section{Cultural Design and Community-Building: Opportunities and Limitations}
Cultural entrenchment is inherent to the growth of large, complex communities. Addressing toxic conflicts early on in peer-production communities can be hindered by different factors. Administrators, moderators and bystanders of peer-production communities may under-prioritize addressing conflicts that can normalize toxicity because of a lack of urgency ~\cite{koehler2018really} or a need for greater context. This means that administrators and moderators may be too late to stop toxic norms from becoming culturally entrenched. However, we believe that it is possible to prevent this from occurring in the first place through informed design.

\begin{figure}[h]
  \centering
  \includegraphics[width=0.75\textwidth]{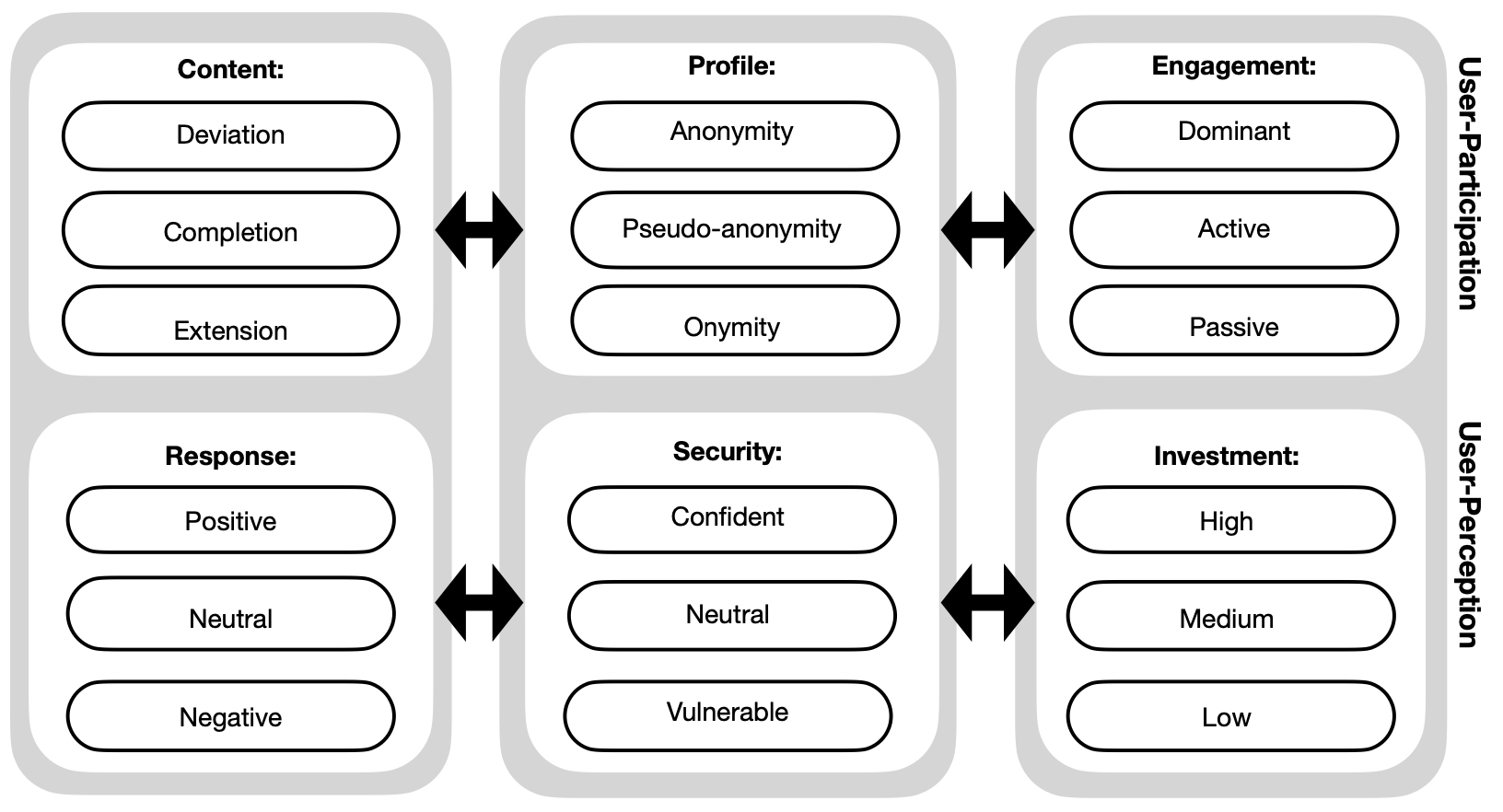}
  \caption{Proposed Framework for Intelligent System Design: The framework consists of six areas of user phenonema. User-participation (top row) refers to quantifiable attributes that identify and document user roles and community structure. Content refers to interactions between users and how much of those interactions deviate from the communication discourse, how much complete or fulfill it, and how much extend the discourse into tangentially relevant new topics. Profile refers to the amount of user information publicly available to the community from complete anonymity, to pseudo-anonymity to full disclosure of personal details (e.g. real name, email, etc.). Engagement refers to a user's role in the community and the proportion of users in dominant contributor roles (e.g. a leader), actively engaged in discussions or passive lurkers. User-perception (bottom row) refers to qualitative feelings and impressions from users in the peer-production community, which may be observable publicly or may be ascertained from user interviews. Response refers to how users respond to content posted in a discussion, whether it is positive, neutral or negative. Security refers to how users feel about their place in the community, whether they are confident with their role, have neutral feelings or feel vulnerable. Investment refers to how emotionally invested users are in their interactions with others, if they are  highly invested (e.g. the topic is of great interest) to medium investment to low investment (e.g. they are unconcerned with how the interaction turns out).}
  \Description{A Proposed Framework for Intelligent System Design.}
\end{figure}

We propose developing a “Handbook for Intelligent System Design” that gathers effective strategies from past community case studies across different domains, in addition to findings from cultural evolution~\cite{zhong2022quantifying, moser2022organizational}, psychology~\cite{cialdini2007descriptive, koehler2018really}, sociology ~\cite{opp2001norms,ziegele2018socially, campbell2009conflict}, and human-computer interaction ~\cite{blackwell2018online, ouvrein2018online, cheng2017anyone}. We suggest the following framework (see Fig. 1) as a foundation for evaluating cultural norms and interactions in peer-production communities. Limitations include domain issues for different communities, differing cultural views from users on what constitutes bullying or harassment, optimal modes for delivering changes and the impact of varying degrees of anonymity. Opportunities include the large number of communities with problems, tools to document and map when and how problems occur, means to observe mitigation attempts and measure how communities feel about themselves.

\bibliographystyle{ACM-Reference-Format}
\bibliography{toxicity-bibliography}

\appendix

\end{document}